# Toward the zero surface tension limit: The granular fingering instability

Xiang Cheng, Lei Xu*, Aaron Patterson, Heinrich M. Jaeger & Sidney R. Nagel

*James Franck Institute and Department of Physics, The University of Chicago, Chicago, IL 60637 USA*

* Present address: Department of Physics, Harvard University, Cambridge, MA 02138, USA.

**The finger-like branching pattern that occurs when a less viscous fluid displaces a more viscous one confined between two parallel plates has been widely studied as a classical example of a mathematically-tractable hydrodynamic instability since the time of Saffman and Taylor[1-3]. Fingering in such Hele-Shaw geometries have been generated not only with Newtonian fluids[4-6] but also with various non-Newtonian fluids[7-9] including fine granular material displaced by gas, liquid, or large grains[10-15]. Here we study a granular Hele-Shaw system to explore whether the absence of cohesive forces in dry granular material can produce an ideal venue for studying the hitherto-unrealizable singular hydrodynamics predicted in the zero-surface-tension limit[2,16-23]. We demonstrate that the grain-gas interface does indeed exhibit fractal structure and sharp cusps associated with finite-time singularities. Above the yield stress, the scaling for the finger width is distinct from that for ordinary fluids, reflecting unique granular properties such as friction-induced dissipation as opposed to viscous damping[24-27]. Despite such differences, the dimension of the global fractal structure and the shape of the singular cusps on the interface agree with the theories based on simple Laplacian growth of conventional fluid fingering in the zero-surface-tension limit[2,16-22]. Our study provides new insights not only on the dynamics of two-phase dense granular flows[24-26] and granular pattern formation[27], but also on the fluid dynamics in the zero-surface-tension limit[2,16].**

Our granular material consists of spherical glass beads with diameters for separate experiments of: $d$=54±10μm, 107±19μm and 359±61μm. The material is loaded into a traditional radial Hele-Shaw cell consisting of two 1.9cm thick circular glass plates of diameter $L$. In most experiments, $L$=50.8cm, but smaller plates with $L$=25.3cm and 17.5cm are also used. The gap between the plates, $b$, is uniform and in different experiments was varied between 0.25mm<$b$<1.14mm. The two plates are maintained at a fixed gap with spacers and clamps around the perimeter. The granular material driven by the displacing gas can flow freely out of the cell boundary. The system was vibrated so that the material between the plates was uniformly dense with a packing fraction near that of random close packing. In order to drive off any humidity, which can cause cohesion between particles, the beads are baked under vacuum before use. We estimated the cohesive force between beads from the angle of the repose of the material[28]. The ratio between the cohesive force and the weight of beads is 6.9% for 54μm beads and 3.5% for 107μm beads. By calculating the surplus surface energy resulting from this cohesive force, we find that the effective surface tension of the granular material is about $10^{-6}$mJ/m$^2$ for both 54μm and 107μm beads, which is 7 orders of magnitude smaller than that of normal fluids such as water (72.8mJ/m$^2$) or ethyl alcohol (22.3mJ/m$^2$).

Pressurized nitrogen, with pressure $\Delta P$ higher than ambient pressure, is blown through a 2mm hole in the center of the bottom plate. The patterns formed are recorded by a high-speed camera (Phantom V7.1) at 2000 frames per second or by a high-resolution still camera. In order to study the dynamics of granular fingering near the yield stress of the granular material, corresponding to a threshold pressure $\Delta P_{th}$, we also conducted our experiment under small constant flow-rate conditions using a syringe pushed by a speed-controlled motor. With a flow rate of 22.9±0.1 ml/s at the smallest gap thickness used, $b$=0.25mm, this covered 0.04% of the entire plate area per second. The patterns grow by stick-slip motion. When the gas pressure is above $\Delta P_{th}$, patterns grow leading to a drop in pressure so that finger growth halts. Then the gas pressure builds up until the fingers resume their expansion. This method forces the pattern to grow under conditions such that the pressure is always close to $\Delta P_{th}$.

Fig.1a and b shows viscous fingering patterns in the granular Hele-Shaw experiments. The shape and dynamics of the pattern depend on the driving pressure, $\Delta P$. When below the yield stress, patterns do not form. Above the yield stress, there are two stages of growth that can be seen by measuring the size of the fastest moving finger. In the early stage, the finger grows at a slow, approximately constant, speed as seen in Fig.1c. In this regime, shown in Fig.1a, one can see a pattern similar to those in fluids. In the late stage, shown in Fig.1b, possibly influenced by the boundary, the growth is accelerated and the patterns grow wildly. In this regime, one sees the merging of fingers and the pinch-off of fingers from the main structure. Such phenomena are rare in fluid fingering. Fig.1c shows that the two stages are connected by a sharp kink at low $\Delta P$ that becomes smoother as the pressure increases. We focus here on the patterns in the early stage.

Fig.2a-c shows the patterns of fingers at different $\Delta P$. The patterns are sharper and more ramified at low pressure and become smoother and more circular at high pressure. This is opposite to what happens in Newtonian fluids where the width of a finger, $W$, depends on the capillary number Ca as

$$W = \pi b \sqrt{1/\mathrm{Ca}} = \pi b \sqrt{\sigma/(\eta V)} \tag{1}$$

where $\sigma$ is the surface tension acting across an interface, $\eta$ is the viscosity of the more viscous fluid (ignoring the viscosity of the less viscous fluid since it is usually much smaller) and $V$ is the local interface velocity[2,5]. The competition between viscous forces, which narrows fingers to a singular point, and surface-tension, which blunts sharp structures in the interface, determines the characteristic length scale of the pattern. Hence, decreasing the driving pressure, which slows down pattern growth, results in more circular patterns in fluid fingering[6].

As in Newtonian fluid fingering, granular fingering grows through tip-splitting instead of side-branching[2]. However, rather than exhibiting an obvious characteristic length scale as found in fluids, the granular fingers widen during growth. To quantify the growth we define the characteristic width of our pattern, $W$, as the width of fingers just before splitting, as shown in lower inset of Fig.2d. We measure $W$ versus the local interface velocity, $V$, at the tip of fingers for systems with different gap thickness, $b$, and plate diameter, $L$, and with different granular material density $\rho$ and size $d$. The results of these studies are shown in the upper inset of Fig.2d. As shown, higher velocities produce wider fingers. The main panel of Fig.2d shows that we can collapse all data onto a single curve with a scaling $W \sim (VLd)^{1/2} \rho^{1/4}$. Comparing with the scaling of classical fluid fingering, Eq.1, the width of granular fingers grows as $V^{1/2}$ rather than decreasing $V^{-1/2}$ and it is independent of the gap thickness $b$. Furthermore, the parameters of the granular particles themselves, the size and density of the beads, enter the scaling.

To quantify global features of patterns at different $\Delta P$, we calculate their fractal dimension by measuring the number of boxes, $N$, needed to cover the entire pattern, as a function of the relative size of the box, $l$. The analyzed patterns are taken from the end of the early stage. We find that the dimension of patterns, $D$, defined as the power of $N \sim l^D$, is limited between that of a circle, $D=2$, and that of a diffusion-limit-aggregation (DLA) pattern[29], $D=1.70\pm0.02$.

In the zero surface-tension limit, viscous fingering patterns are predicted to have the same fractal dimension as DLA[2,7,17,18]. As we decrease the driving pressure towards the threshold pressure, $\Delta P_{th}$, we find that $D$ approaches the exponent characteristic of DLA. This suggests that granular fingering may only show the zero surface-tension-fluid behavior or the anticipated singularity near the threshold pressure, $\Delta P_{th}$. To investigate this regime in detail, we conducted a constant flow-rate experiment as described above. The fractal dimension of the resulting pattern ($D=1.68\pm0.02$) indeed overlaps with that of DLA as seen in Fig.3a.

We can also focus on a small region along the interface and observe cusp formation as seen in upper inset of Fig.3b. These cusp tips can be sharp down to the grain level. Cusp formation occurs just before each finger starts to grow. These cusps are sharp and different from any structures found in Newtonian fluids where the finger tips are always round due to surface-tension[4-6]. Theory suggests[2,16,20-22,30] that the shape of the profile just as it becomes singular should scale as

$$y=A(x-x_0-u)^{0.5} (x-x_0+2u) \qquad (2)$$

where $A$ is a free parameter, and $u$ is the location of the finger tip at time $t$ (up to a $x_0$ shift). The $y$-axis has been shifted so that the cusp at the singular time $t_c$ is located at $x=x_0$ and $y=0$. The time evolution of $u$ is dictated by $u=B*\text{sqrt}(t_c-t)$, where $B$ is a prefactor[30]. Therefore, the velocity of the finger tip, $v$, is given by $v=du/dt=-B/2\text{sqrt}(t_c-t)$. In terms of $u$, $v=-B^2/2u$. The lower inset to Fig.3b shows the time evolution of an individual cusp. Due to the stick-slip motion of the interface and also due to the rapid movement of the finger tips, we cannot reliably resolve the -1/2 time scaling of *v*. However, we can collapse the profiles at *u*=0, the point where they are sharpest, onto a single master curve: $y=A(x-x_0)^\beta$ with $\beta$=1.43±0.20 as shown in Fig.3b. This scaling is consistent with Eq.2, $\beta$=3/2, at the instant of cusp formation.

Because of the absence of surface-tension and the irrelevance of thermal energy in granular material, the interfacial patterns formed in our system do not relax when we turn off the driving pressure. Without mechanical perturbation, patterns such as those shown in Fig.2 retain their sharp structure. Hence, our system provides an opportunity to study the steady-state structure of the patterns. This is in contrast to other systems where surface tension and thermal diffusion always smooth out the interface[31].

In conclusion, we have studied the dynamics of fingering in the Hele-Shaw geometry when air is injected into dry granular material. Just above the yield stress, when the granular material just starts to flow, the dynamics produces singular features as expected for a fluid with zero surface-tension: (i) cusps in the interface which collapse onto a theoretically predicted master curve and (ii) a fractal dimension similar to that found for DLA. Theoretically, it has been known that fluid fingering in the zero surface-tension limit would lead to singular cusp structure locally and DLA fractal geometry globally[2,16-22]. However, the connection between the local singular dynamics and global fractal shape has been far from clear despite extensive study. Our results provide experimental evidence for the coexistence of these two singular features and shows how the local cusps evolve into a global fractal. Moreover, we have identified the scaling of a granular fingering instability which is in contrast to what is found in ordinary fluids. It is intriguing to note that the particle diameter, *d*, enters the scaling as an important parameter. Without surface tension to provide a scale to regularize the singularity, the grain diameter may provide that scale.

We thank E. Bettelheim, E. Corwin, S.Y. Lee, J. Royer, P. Wiegmann and L.N. Zou for fruitful discussions. This work was supported by the MRSEC program of the NSF under DMR-0213745.

Correspondence and requests for materials should be addressed to X.C. (xcheng@uchicago.edu).

Fig.1: Granular fingering. Finger patterns during (a) early stage, $t$=0.08s, and (b) late stage, $t$=0.168s, of growth. The grey regions in the images are still-undisturbed glass beads and the black regions are empty areas where the gas has displaced the beads. The data were obtained with beads of diameter $d$=107μm, plate separation $b$=0.51mm, system diameter $L$=50.8cm and fixed gas overpressure $\Delta P$=0.48atm. The scale bar is 3.0cm. $R$ is defined as the radius of the longest finger measured from the central hole, as shown by the dotted circle in (a). (c) $R$ versus renormalized time $t/t_0$. The early and late stage are separated at $t=t_0$. Symbols from left (black square) to the right (pink diamond) in the late stage are $\Delta P$=0.15atm, 0.20atm, 0.34atm, 0.68atm and 1.02atm respectively. The horizontal dotted line marks the boundary of the cell. Inset: The time $t_0$ as a function of $\Delta P$. It appears to diverge at $\Delta P_{th}$ (vertical dashed line) where patterns no longer grow.

Fig.2: Scaling of the granular fingering instability. (a)-(c) Patterns at different overpressure $\Delta P$ with $d$=54µm, $b$=0.25mm and $L$=50.8cm. $\Delta P$= 0.10atm (a), 0.20atm (b), 1.36atm (c). Scale bars are 1.5cm. Tips of fingers at low pressure are much finer and sharper than those at high pressure. Similarly, the pattern is more like DLA near $\Delta P_{th}$=0.09atm. (c) Scaling of finger width, $W$, versus $VLd\rho^{1/2}$. Black line is the best power law fit, with slope 0.51±0.01. Lower inset: Growth and tip-splitting of fingers at $\Delta P$=0.15atm with $d$=107µm, $b$=0.51mm and $L$=50.8cm. Scale bar is 0.8cm. Time between two images is 0.028s. As shown, $W$ is defined as the finger width just before it splits. Notice the sharp cusp shown in the second picture during the growth of a finger. Upper inset: Finger $W$ versus tip velocity $V$. Data shown are for $d$=107µm glass beads ($\rho$=2.6g/cm$^3$) with $L$=50.8cm and $b$=0.254mm (N1), $L$=50.8cm and $b$=0.508mm (N2), $L$=50.8cm and $b$=0.762mm (N3), $L$=50.8cm and $b$=1.143mm (N4), $L$=25.3cm and $b$=0.508mm (N8), $L$=17.5cm and $b$=0.508mm (N9); for $d$=54µm glass beads with $L$=50.8cm and $b$=0.254mm (N6); for $d$=359µm glass beads with $L$=50.8cm and $b$=1.143mm (N7) and for $d$=130µm copper beads ($\rho$=8.4g/cm$^3$) with $L$=50.8cm and $b$=0.508mm (N5).

Fig.3: Singular dynamics of granular fingering. (a) Comparison between a granular fingering pattern near $\Delta P_{th}$ and DLA. Shown are box counting results for a pattern generated under constant flow rate conditions ($d$=54µm and $b$=0.25mm) and for a DLA pattern generated numerically with the same resolution as that of the granular pattern. $N$ is the number of boxes of side length $l$ needed to cover a pattern. Solid line indicates a power law with slope 1.70. Inset: Local slope of the box counting results, corresponding to the local fractal dimension, $D_{local}$, as computed by a linear least squares fit over the sliding interval $\Delta\log(l)$=0.45. The local slope of our pattern at constant flow rate matches that of the DLA pattern except for the smallest box sizes. The dimension of our DLA pattern is a little smaller than the literature value[29] 1.70±0.02 due to the finite resolution. (b) Cusp structure in the interface. Upper inset: Three representative images of the air-sand interface for $d$=107µm, $b$=0.51mm and $L$=50.8cm. Scale bars are 1.5mm. Individual grains are seen in the pictures. Lower inset: Time evolution of an individual cusp. Scale bar is 0.5mm. Main panel: Scaling of the cusp profiles. Data shown are combined from different experiments, from different cusps in the same experiment, and from left and right arms of each cusp. Vertical and horizontal axes are in units of particle size, $d$, and are normalized according Eq.2. Note that we follow the theoretical papers[20-22,30] which define the x-axis as the growth direction so that the cusp tip coincides with y=0. The line shows a power law with slope 3/2. We identified the cusp structure before the growth of almost every finger for small particles (54µm and 107µm) near $\Delta P_{th}$.

**Figure 1:**

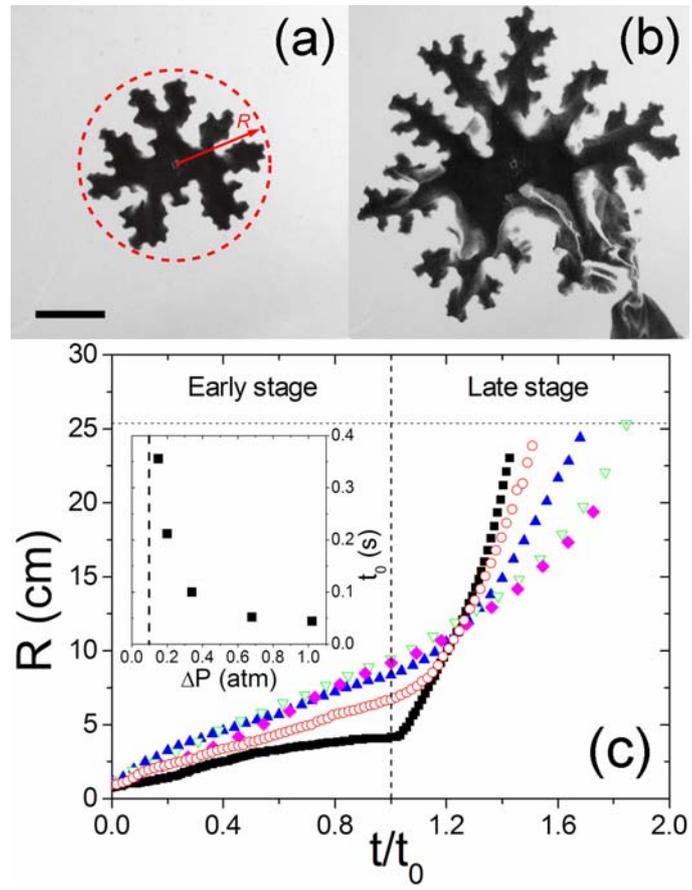

**Figure 2:**

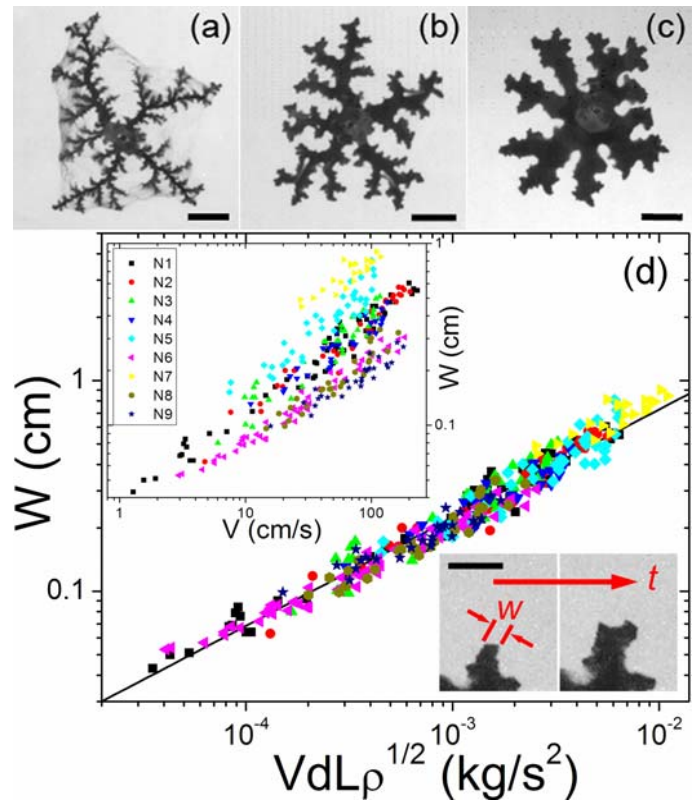

**Figure 3:**

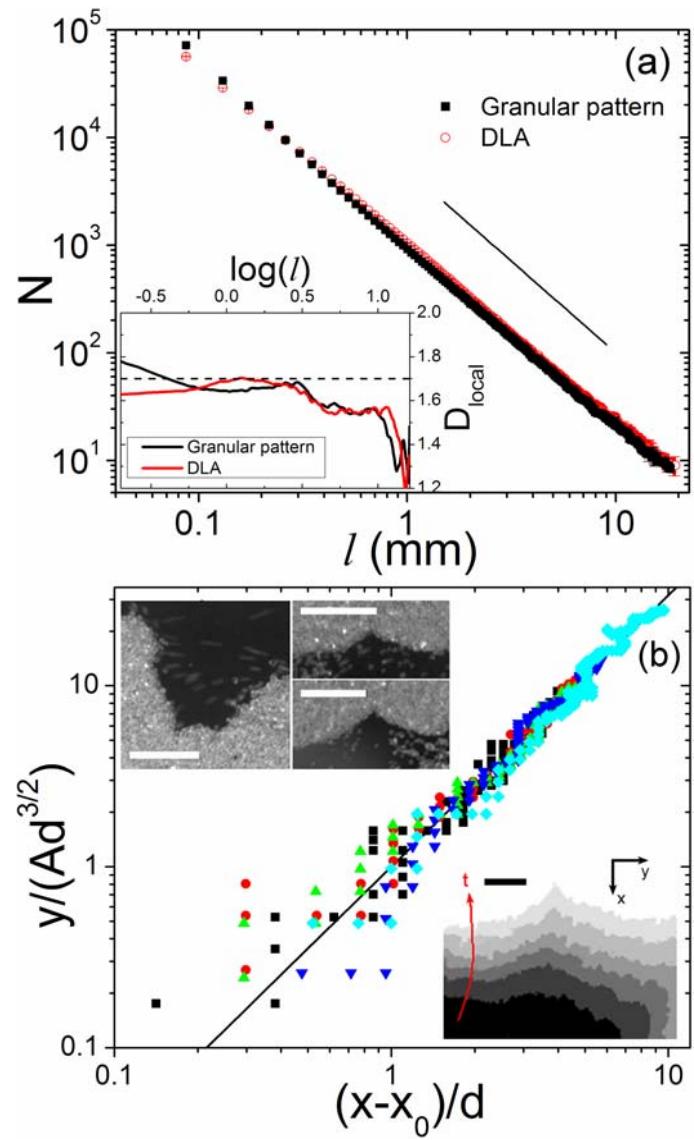